\documentclass{article}
\usepackage[utf8]{inputenc}
\usepackage{hyperref}
\usepackage{graphicx}

\title{Bayesian Modeling of Marketing Attribution}
\author{Ritwik Sinha\thanks{Adobe Research, 345 Park Ave, San Jose, CA 95110} \and  David Arbour \thanks{Adobe Research, 345 Park Ave, San Jose, CA 95110} \and  Aahlad Manas Puli\thanks{New York University,  60 Fifth Ave, New York, NY 10011} }

\begin{document}

\maketitle
\begin{abstract}
In a multi-channel marketing world, the purchase decision journey encounters many interactions (e.g., email, mobile notifications, display advertising, social media, and so on). These impressions have direct (main effects), as well as interactive influence on the final decision of the customer. To maximize conversions, a marketer needs to understand how each of these marketing efforts individually and collectively affect the customer's final decision. This insight will help her optimize the advertising budget over interacting marketing channels. This problem of interpreting the influence of various marketing channels to the customer's decision process is called marketing attribution. We propose a Bayesian model of marketing attribution that captures established modes of action of advertisements, including the direct effect of the ad, decay of the ad effect, interaction between ads, and customer heterogeneity. Our model allows us to incorporate information from customer's features and provides usable error bounds for parameters of interest, like the ad effect or the half-life of an ad. We apply our model on a real-world dataset and evaluate its performance against alternatives in simulations. 
\end{abstract}

\section{Introduction}
Digital marketing has given the marketer greater access to potential customers as well an ability to observe every interactions between the brand and the customer. These interactions may further happen across a gamut of different marketing channels (for example, Email, mobile, display advertising, social media, and so on) over several days or months. This access to digital marketing data has allowed a marketer to assign credit for each conversion event to various marketing channels. This is called the multi-channel marketing attribution (attribution, for short) problem.

These marketing interactions between the customer and the brand have direct and synergistic influence on the final decision of the customer. To maximize conversions, a marketer needs to understand how each of these marketing efforts individually and collectively affect the customer's final decision and accordingly, to optimize her advertising budget over all marketing channels. Traditional attribution models assign the influence to each marketing channels in a rule-based manner, which are often non-intuitive. This motivates the need for finding the influence of each channel in an algorithmic fashion. Although there have been multiple algorithmic models, they continue to be simplistic, failing to capture all the mechanisms in which a customers arrive at their purchase decisions. 

Current methods for attributing conversions to marketing channels range from simple rules to sophisticated algorithms. The simplistic methods or the single-channel attribution models give the whole credit of the conversions to either the first step (First touch attribution), the last step (Last touch attribution), or give equal weights to all touches (Linear touch attribution) of a customer's purchase journey. Such approaches neglect to contrast converting with non-converting paths. Data driven attribution models go further. In this literature, the solution to the problem of marketing attribution has two steps. First, estimate the behavior of a customer when exposed to variety of marketing channels. Second, using these probabilistic models of customer behavior, the estimates from the first stage are interpreted as the influence of each channel to finally come up with the estimated attributions of the various marketing channels. In this work, we improve the first step, and rely on the state of the art for the second step of the solution~\cite{Yadagiri2015ANA, sinha2014estimating}. That brings us to the following problem statement.

In this work, we contribute in the following ways. We propose a model that can capture the synergistic effects between marketing channels to arrive at a flexible probabilistic customer behavior model for attribution. In particular, our model estimates marketing characteristics like the following: (1) The direct effect of a marketing channel interaction, (2) The decay of this effect, which allows us to know the half life of an ad, (3) The interaction or synergistic effects between ads, (4) The ability to model customer heterogeneity (impulsive vs. careful buyers), (5) The ability to control for observed features of a customer, (6) Generate error estimates for all estimated parameters. We test our model on a simulated dataset (with known values of these parameters) and a real-world marketing dataset from a digital marketing brand. 

\section{Related Work}

Marketing Attribution is well studied in the the digital marketing industry. It started with simple rule based models providing insights into the possible role played by different marketing interactions towards some marketing outcome. There are no parameters to estimate in these models, rather they are frequency based estimates. A few examples are \textit{first-touch, last-touch}, \textit{equal}, and \textit{time decay}. First touch and last touch models assign all the credit to the first and last touches respectively. While equal assigns equal credit, and time decay assigns diminishing credit to all touches \cite{Attribut50:online, Overview70:online}. 

The growth in digital marketing and the ease of data collection has opened up the possibility of building data-driven models for attribution. The first data driven approach to solving this problem \cite{shao2011data} involved frequency based methods to compute the relative credits for different marketing channels. A causal framework for attribution was proposed in \cite{dalessandro2012causally}. This approach however requires untestable assumptions including no unobserved confounders and positivity. An econometric model that computes the incremental effect of a marketing interaction is proposed in \cite{sinha2014estimating}, requiring predictive models for conversion likelihood. Shapley value \cite{shapley1953stochastic} based attribution models have been proposed \cite{Yadagiri2015ANA, singal2019shapley}, but these do not scale to the typically high cardinality of marketing channels (these approaches are exponential in the cardinality of the dimension). More recent work along these lines include scalable Shapley value based methods \cite{sinha2020attribution}, unfortunately, this approach does not use the non-converting paths within the modeling step. 

Markov chain based attribution models have been proposed \cite{anderl2014mapping, kakalejvcik2018multichannel}. While appealing, these also suffer from limitations in how big the cardinality of the dimension can be (being quadratic in the cardinality for the simplest model). Models that rely on deep neural nets, including attention based models \cite{ren2018learning} and recurrent neural nets \cite{du2019causally}, have also been studied. However, since these models require computing the parameters of a deep neural network from a large number of customer journeys, these cannot be computed at scale. Further, none of these models estimate parameters of interest like decay in the effect of the ad or customer heterogeneity. This motivates the need for a theoretically well motivated  attribution model that captures the marketing characteristics of interest.


A recent line of work~\cite{jin2017bayesian} has exploited the Bayesian treatment of the marketing media mix problem. They target are carry-over and shape effects; and consequently the authors focus on time-series modelling of the delayed customer response for \textit{each} separate ad channel. To achieve this, they posit a Bayesian model using a non-parametric function and then approximate it. Their techniques do not help account for the interactions between two ad channels. Also, this model only captures aggregate effects, as they operate on the time series of exposures and outcomes. Our model is more closely related to~\cite{Yadagiri2015ANA} but improves upon it by capturing the influence of time-difference and quantifying uncertainty.

\section{Methods}
We propose a generalized framework of models that subsumes existing attribution techniques with a likelihood-based model. Some of the position based methods exist as special cases of our framework. Moreover, simple design choices allow for the incorporation time-dependence and interaction into the models in our framework.
Our model is designed with parameters that correspond to many real-world attributes of the marketing pipeline. Moreover, the Bayesian formulation provides us with a non-trivial model validation measure in the likelihood score. Unlike existing models, we explicitly model the following elements of how customers interact with the marketing system, and do so with \textbf{interpretable} parameters. First, we capture customer fatigue and heterogeneity. Second, estimate the decay of each ad interaction with respect to time. Finally, we model the interaction between different marketing channels. 


\subsection{Workflow of the Model}
Our model fits into the marketing pipeline in between the data-collection and ad-assignment optimization stage.
Given data from recent marketing history of a product, our model provides distributions over attributions that reflect uncertainty inherent in the data. Moreover, our model allows for distributions over the interaction effects among the different marketing channels. These advantages allow the marketer to calibrate their faith in their ad campaigns.
\begin{figure}[ht]
\centering
\includegraphics[width=0.8\textwidth]{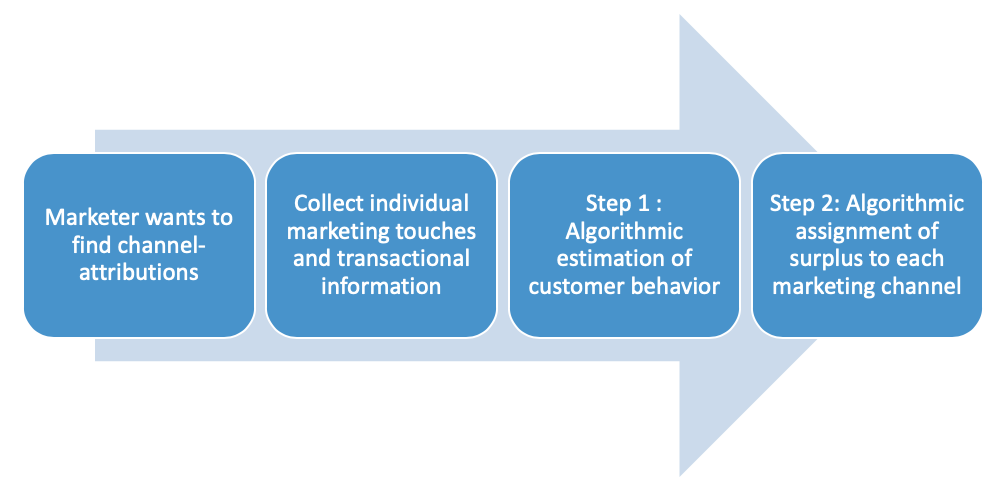}
\caption{The workflow used by the marketer to optimize marketing spend across channels and campaigns. Starting with the business question, data collection, estimation of the behavior model, and credit assignment.}
\end{figure}

\subsection{Model}

We model the likelihood of a customer to make a purchase at each time of interaction as a function of the prior treatment times and channels. We use the following parameters in our model, each of which corresponds a distinct real-world quantity. Let $r_{i, t_k}$ denote the random variable of whether customer $i$ makes a purchase at time $t_k$. Also, let $a_j$ denote the treatment (marketing channel) at time $t_j$. The first parameter of interest is $\mu$, which captures the customer-independent baseline chance of purchase. The term $b_i$ reflects customer-heterogeneity of the $i$th customer, this is assumed to be normally distributed with $0$-mean. Next, we define $\beta_a, \lambda_a$ as the channel-specific base magnitude and decay parameter respectively. The parameter $\gamma$ controls the strength and direction of interactive effects.

In our model, $g$ is some link function that specifies the class of predictions; e.g., the logit link for a binary response.

\begin{eqnarray}
\label{eqn:main}
p(r_{t_k} = 1 | a_1, t_1,\cdots , a_k,t_k) &=& g(\mu + b_i \nonumber  \\
& + &  \sum_i \beta_{a_i}\lambda_{a_i}^{t_i} + \sum_{i\not = j} \gamma\beta_{a_i}\beta_{a_j}\lambda_{a_i}^{t_i} \lambda_{a_j}^{t_j} )
\end{eqnarray}
    
Furthermore, an advantage of this likelihood formulation is that we can further incorporate elements like conditioning on previous sales and other observed characteristics of a customer by readily including new covariates in this model. 

\begin{figure}[ht]
\centering
\includegraphics[width=0.7\textwidth]{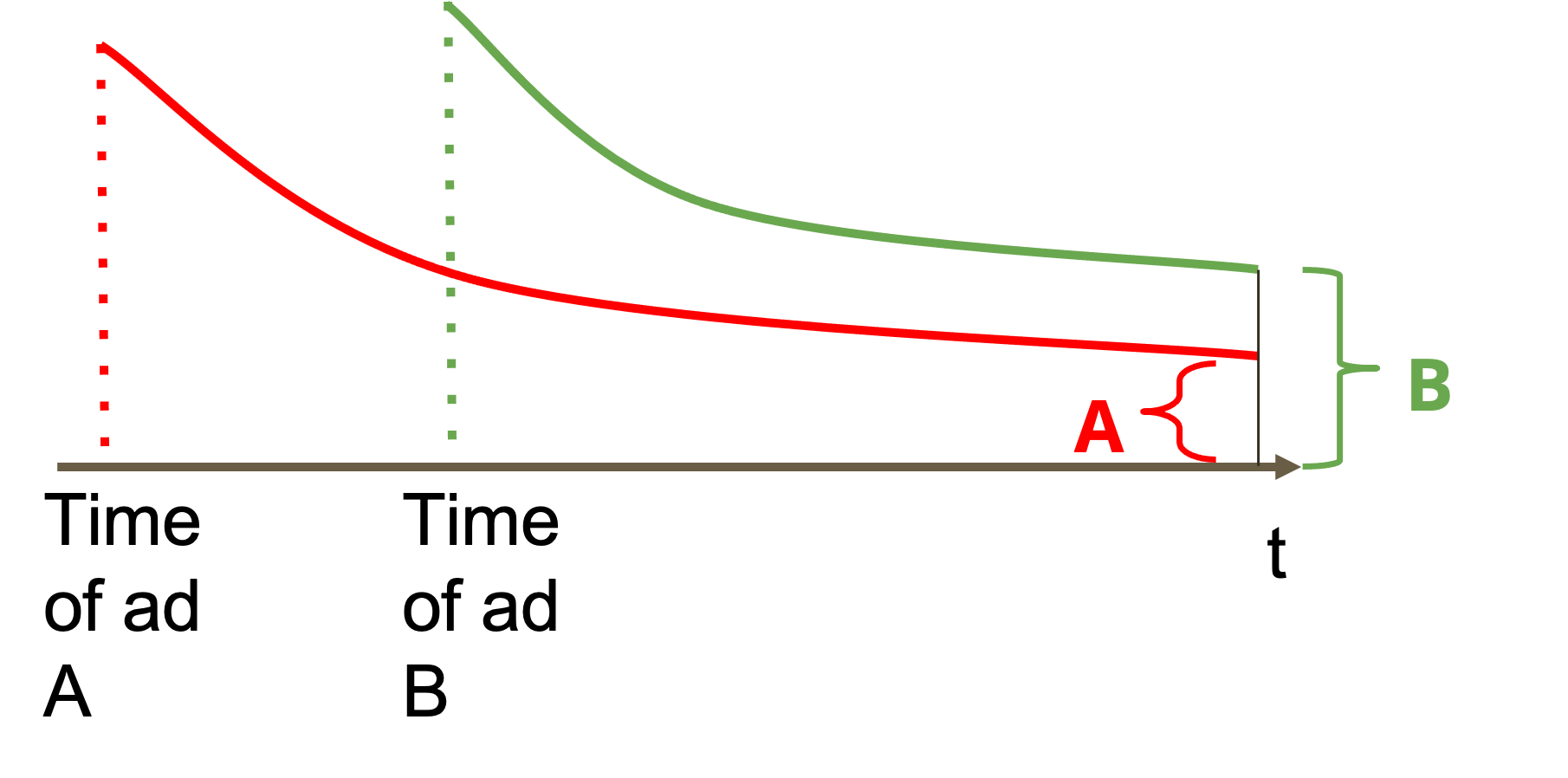}
\caption{This figure describes the main idea behind the underlying model. Along the timeline are displayed the times of the two ads $A$ and $B$, the height of the lines denote the main effect of these ads, as time passes, the effect decays, this is denoted by the two curves. At time $t$, the remaining effect of the two ads is denoted by the height of the curves at this point. In addition to the main effects, there may be an interaction between all the different channels that this person has been exposed to. }
\end{figure}

\subsection{Bayesian Solution to the Model}

We propose a Bayesian solution to the likelihood defined in (Equation \ref{eqn:main}). We specify the following priors on the parameters, with wide and flat priors encoding minimal or no prior information into the model.
\def\cN{\mathcal{N}}
Customer-heterogeneity is modelled as a random-effect $b_i$ with variance regularized to be small. We assume $b_i \sim \cN(0,\sigma_b^2),$ where $\sigma_b\sim \exp(0.5)$. We assume a wide non-negative prior on the base magnitudes $\beta_a\sim \exp(10)$. The sign-restriction on $\beta$s reflects the knowledge that all ads have a non-negative effect when they occur without any interactions. Next, we assume wide priors on the interaction term and the baseline $\gamma \sim \cN(0,10),\quad \mu \sim \cN(0,10)$. Finally, a flat positive prior is assumed on the decay parameters $\lambda_a\sim Unif(0,1)$. This ensures that the decay always takes the main effect of the ad to $0$ as time passes. 

\section{Results}

We present the results of our model on a simulated dataset and one a real-world dataset. We first describe the results of our model on the simulated dataset. 

\begin{figure}[ht]
\centering
 \includegraphics[width=0.7\textwidth]{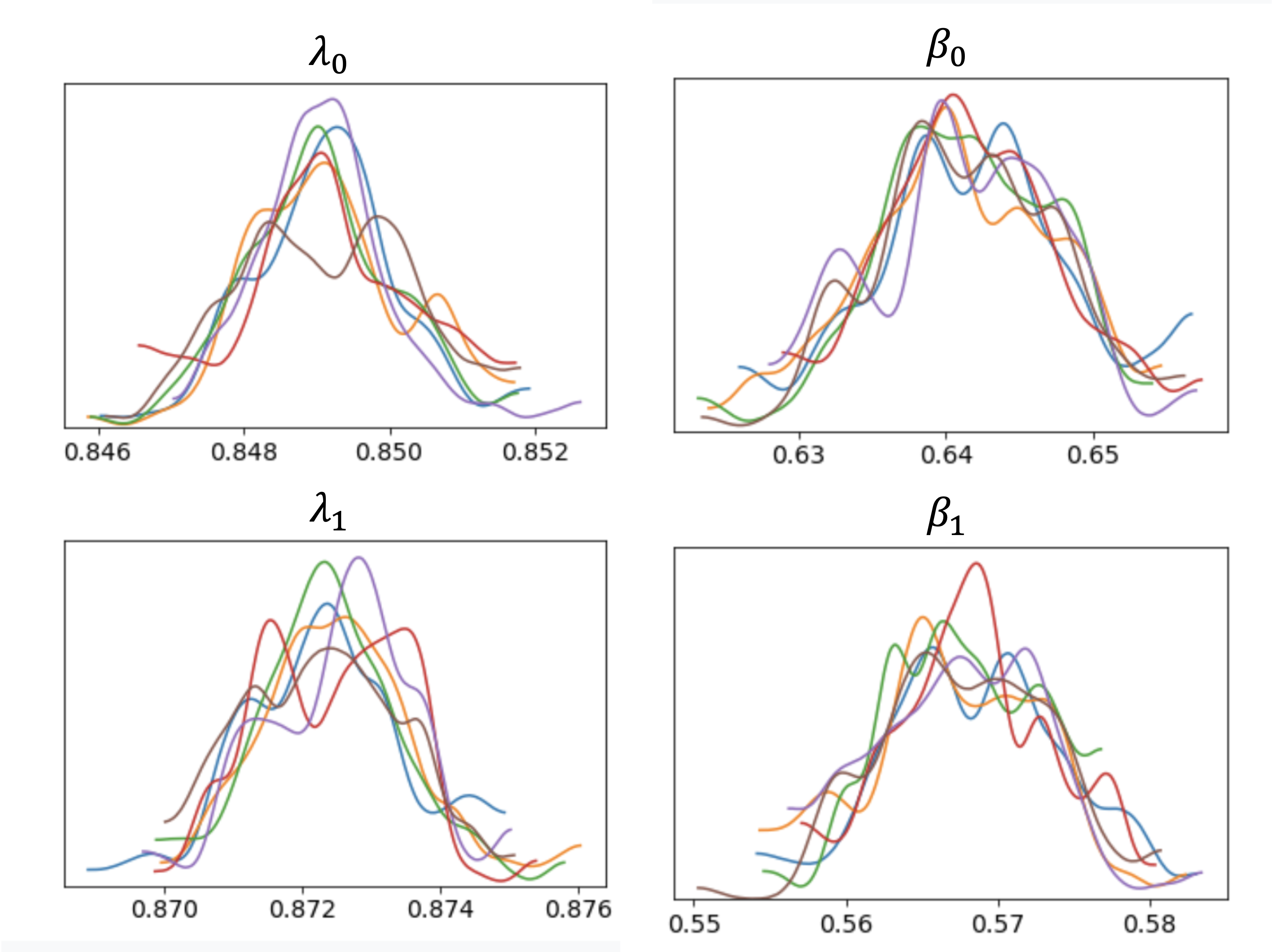}
\caption{Sharp estimates for posteriors of parameters of 2 different touches.}\label{fig:sim-results}
\end{figure}

\subsection{Results on Simulated Data}

We chose to work with a regression setup for the simulated experiment. That is,  the link function $g$ is assumed to be the identity function. In our simulation, we do not incorporate random effects and baseline parameters. The synthetic data was generated as follows. We generate a sample of $10,000$ customer journeys. From a set of $5$ channels or interaction types, we sample $10$ actions, with equal probability and with replacement. To simulate the time between touches, we assume that the inter event time follows an exponential distribution. In other words, $ t_k - t_{k-1} \sim \exp(1)$. The parameters $\beta \sim U(0,1)$ and $\gamma\sim \cN(0,1)$. Finally, we assume that $\lambda\sim \beta(1,1)$, which ensures that the decay is always to zero, and the treatment effect does not explode. The outcome $y$ is generated according to our model with the parameters sampled as above.

We estimated the posterior through Markov Chain Monte Carlo (MCMC) methods implemented in the STAN framework~\cite{carpenter2017stan}. Over multiple runs, we observed that the posteriors were sharp, the chains mixed well and the parameters were recovered well. See Figure~\ref{fig:sim-results} for a few posteriors of $\beta$ and $\lambda$ corresponding to channels $\{0,1\}$. Notice that almost all mass is concentrated within an interval of width $0.02$ (with less than $4\%$ error). This provides us confidence that our strategy can recover the true underlying parameters from such a non-linear regression problem. 

\subsection{Estimates on Real Data}

We apply our approach to a web analytics dataset from a major company from the travel and experience industry. This data is collected using Adobe Analytics. Adobe Analytics provides marketers the most comprehensive set of tools to measure and track all aspects of usage of an organization's website. The data is from the months of September and October of 2013. This data amounts to about 3.6 TB of storage and 2 Billion page views from about 26 Million unique visitors.

\begin{figure}[ht]
\centering
\includegraphics[width=0.7\textwidth]{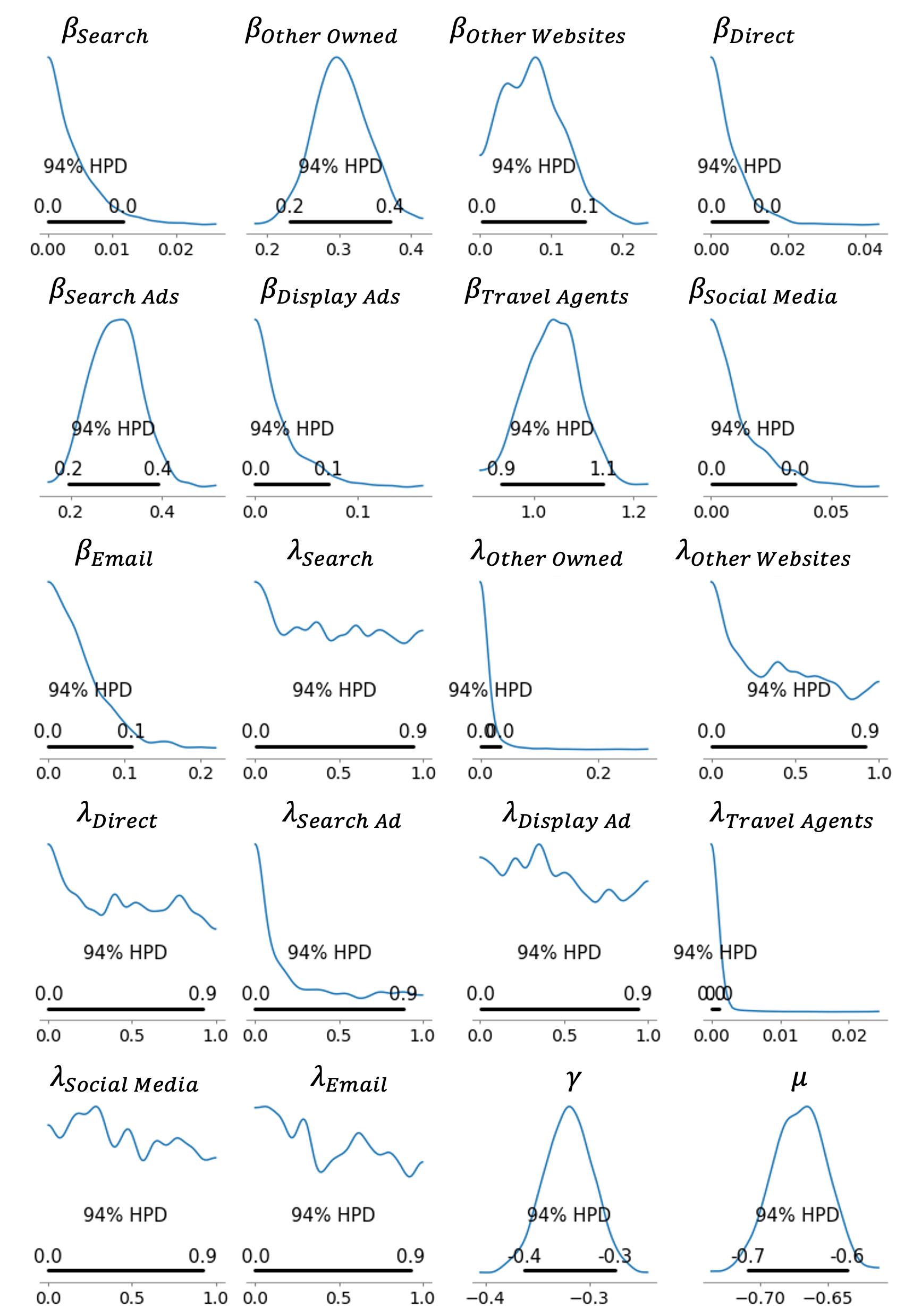}
\caption{Estimates for posteriors of parameters of the model}
\label{fig:real-params}
\end{figure}

To enable quicker testing of our proposed technology, we performed a stratified sampling of our data. We sample about 5 million unique visitors with the following restrictions, all of these visitors have visited the web property during the last two weeks of the data window, and about $300,000$ of them have made a purchase on the website in the last two weeks of October. For all such individuals, we then cover two months of history. The entire data amounted to 5 Million interactions. The visitors could arrive from one of 9 marketing channels. 

We restricted our data to people with at most 5 touch events. Further we sub-sampled customers with no sales to obtain a balanced ratio of about 70 to 30, sale to no-sale. Further, we sampled 5,000, which is around 10,000 events, upon which the parameter estimation was done. We computed the attribution of a channel as the sum over all customers of differences of the probabilities of sale with and without an ad channel. The focus for the real experiment was predicting the occurrence of a sale, meaning we chose $g$ to be a the sigmoid function.

\begin{figure}[ht]
\centering
\includegraphics[width=0.7\textwidth]{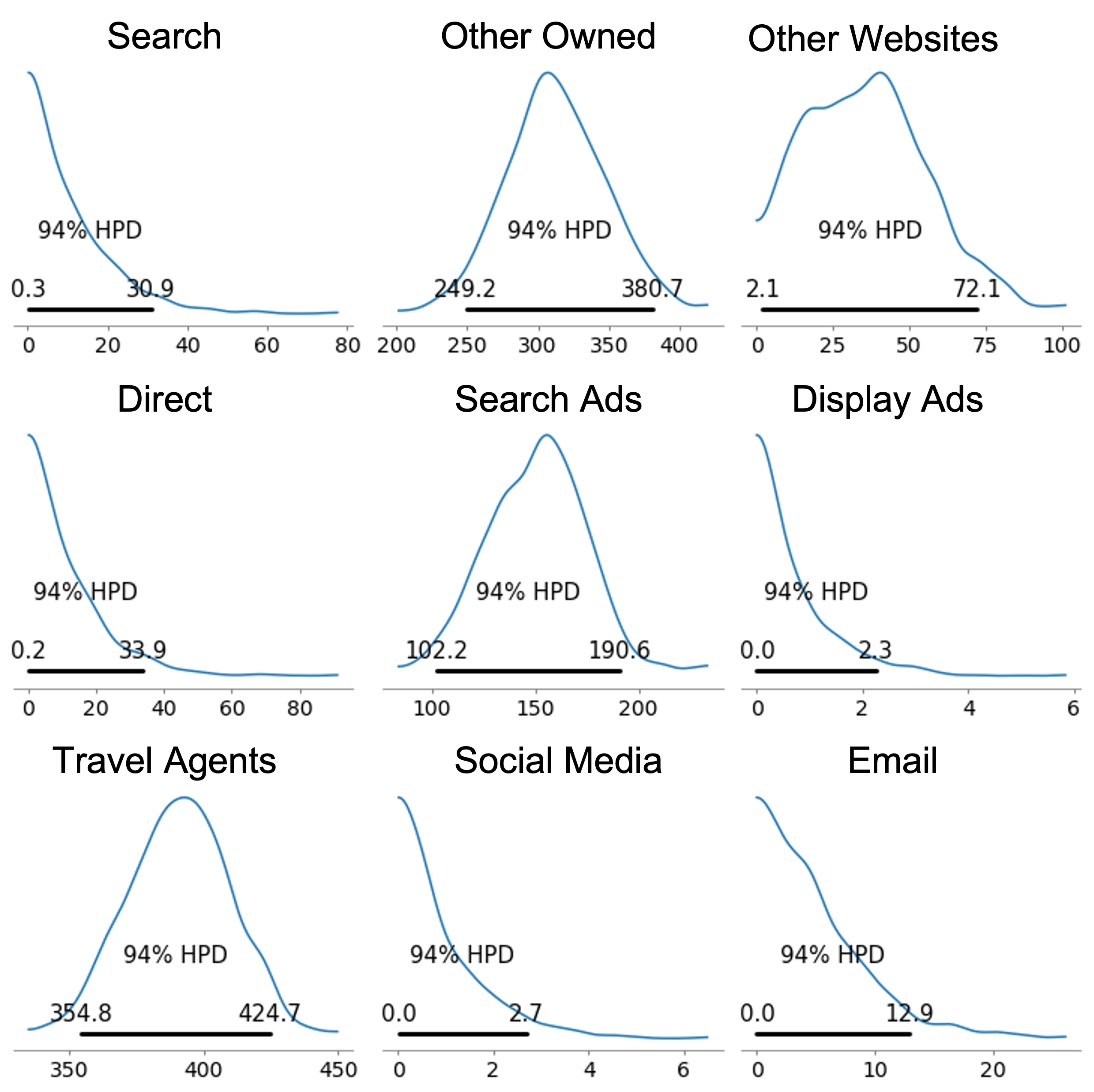}
\caption{Attributions of all touches}
\label{fig:real-attr}
\end{figure}

We plot the estimated posteriors for the parameters and the attributions in Figures~\ref{fig:real-params} and \ref{fig:real-attr} respectively. We see that intuitive phenomena are recovered by examining the sharp posteriors (bottom of Figure~\ref{fig:real-params}) for the parameters $\gamma$ and $\mu$. One interesting thing to note from Figure~\ref{fig:real-params} is as follows. Whenever the $\beta$ for a particular channel is estimated to be close to $0$, the corresponding $\lambda$ has a flat uninformative posterior, for example, $\beta_{Search}$ and $\lambda_{Search}$. This is reasonable given that we are multiplying some power of $\lambda$ by a factor close to $0$. On the other hand, whenever $\beta$ for a parameter is estimated to be distinct from $0$, we observe that the corresponding $\lambda$ is close to $0$, for example, $\beta_{Search Ads}$ and $\lambda_{Search Ads}$. This suggests that, for this dataset, the effect of the ads tend to decay very fast. That is, customer tend to forget the ad impression very rapidly. 

The negative estimate of the interaction term captures the concept of customer fatigue; where a lot of ads ($>20$ in this case) served within a short period of time results in the interactive effects adversely affecting the chances of sale.
The \textit{mu\_bin} term being a large negative quantity reflects that the fact that without any ads served, the chances of purchase are low in general. From Figure~\ref{fig:real-attr} we see that the largest number of orders are attributed to Travel Agents, followed by Other Owned websites, and Search Ads. 

\section{Conclusion}
Existing attribution techniques are insufficient because of two reasons. Rule-based techniques are not flexible and score-based techniques ignore time or order information. By utilizing parametric functions of time-differences between actions and likelihood-based estimation, our framework addresses these problems. Additionally, the framework provides interpretable parameters that are insightful even when the true model is misspecified. Further more, an advantage of the sale-likelihood formulation is that we can further incorporate elements like conditioning on previous sales and other observed characteristics of a customer. A natural area of extension for this work is to capture the situation where visitors are may use multiple devices to interact digitally with the brand, this will mean that the digital marketing data may record these devices as separate customer journeys. Combining these separate journeys probabilistically into a single stitched journey has been explored \cite{saha2015probabilistic}, but incorporating these ideas into the attribution models promises advantages, but presents challenges.

\bibliographystyle{unsrt}
\bibliography{references}
\end{document}